\documentclass[useAMS,usenatbib]{mn2e}
\pdfoutput=1
\usepackage{graphicx}
\usepackage{color}

\title[On the Stability of Tidal Streams]{On the Stability of Tidal Streams}
\author[Aurel Schneider and Ben Moore]{
Aurel Schneider and
Ben Moore \\
{Institute for Theoretical Physics, University of Zurich, Zurich, Switzerland;}\\
{aurel@physik.uzh.ch; moore@physik.uzh.ch}  \\
}

\begin{document}

%\date{Accepted Year Month Day. Received Year Month Day; in original form Year Month Day}

\pagerange{\pageref{firstpage}--\pageref{lastpage}} \pubyear{2011}

\label{firstpage}
\maketitle

\begin{abstract}
We explore the stability of tidal streams to perturbations, motivated by recent claims that the clumpy structure of the stellar streams surrounding the globular cluster Palomar 5 are the result of gravitational instability. We calculate the Jeans length of tidal streams by treating them as a thin expanding cylinder of collisionless matter. We also find a general relation between the density and the velocity dispersion inside a stream, which is used to determine the longitudinal Jeans criterion.
Our analytic results are checked by following the time evolution of the phase space density within streams using numerical simulations. 
We conclude that tidal streams within our galactic halo are stable on all length scales and over all timescales. 
\end{abstract}


\section{Introduction}

Tidal streams are a widespread phenomenon in astrophysics, emerging from star clusters \citep{Grillmair1995}, dark matter subhalos \citep{Diemand2008} or satellites of galaxies \citep{Ibata1994}. It is present on all scales from galaxy clusters \citep{Calcaneo-Roldan2000} down to the the very smallest dark matter substructures \citep{Schneider2010}. The gravitationally unbound material forms spectacular long streams that trace the past and future orbit of the host system.

A wide class of tidal streams can be treated as collisionless systems, since they are dominated by stars or dark matter particles - the local relaxation time within the stream is much longer than the age of the Universe. Some streams contain gaseous material and are much more complicated to understand. For example, the oldest example of a 'stream' is the spectacular Magellanic HI stream, trailing well over 100 degrees behind the Magellanic Clouds. Initially modelled as a tidal mass loss feature from the Large Magellanic Clouds (\citet{Lin1977}), an alternative explanation is that it resulted from a more complex gravitational interaction between the Large and the Small Magellanic Cloud prior to their infall in the Milky Way potential \citep{Besla2010}. However, it may also be the case that this feature is purely hydrodynamical in origin since it contains no stars \citep{Moore1994}.

Galaxy mergers often create spectacular tidal tails that are somewhat different from the streams we consider in this paper. These streams are rapidly and violently created and they can contain dwarf galaxies aligned along the tails. \citet{Barnes1992} carried out simulations of galaxy mergers and found collapsed objects populating the stellar tails. Therefore they proposed collisionless collapse as the creation mechanism of tidal dwarf galaxies. However, \citet{Wetzstein2007} identified these collapsed objects as numerical artefacts due to insufficient resolution. They rather found that it's the gaseous part of the streams that triggers the collapse which leads to tidal dwarf galaxies.

The dynamics of tidal streams from star clusters and dwarf galaxies in our own halo have been extensively studied to constrain the mass and shape of the Galactic potential \citep{Johnston1999,Law2009}, alternative gravity models \citep{Read2005} as well as the orbital history of the satellites \citep{Kallivayalil2006,Lux2010}. However, the detailed evolution of the internal phase space structure of streams has received less attention \citep{Helmi1999,Eyre2010}.

Simulations, as well as observational data, show variations in the width and the internal structure of tidal streams. Likewise the density along a stream can vary considerably, the most prominent example being the symmetric streams originating from the globular cluster Palomar 5 with its equally spaced density clumps \citep{Odenkirchen2001, Odenkirchen2002}. There are different explanations for these clumps such as disc shocking \citep{Dehnen2004}, effects due to the dark matter substructures \citep{Mayer2002,Yoon2010} or epicyclic motions in the stellar orbits \citep{Kuepper2008,Just2009}.

Another interpretation was given recently by Quillen and Comparetta \citep{Quillen2010,Comparetta2010}, who argued that clumps in streams are the result of longitudinal Jeans instabilities. In their model they describe a tidal stream as an extended static cylinder of stars and they use the results of \citet{Fridman1984}, that infinitely extended cylinders are gravitationally unstable. With an estimated relation for the velocity dispersion and the linear density in the stream, Quillen \& Comparetta find a longitudinal Jeans length of several times the stream width. Comparing their results to the observations of Palomar 5, they find agreement between the distance between clumps in the streams and their fastest growing mode of the gravitational instability.

However, their model of a static cylinder does not take into account the expansion that happens due to the diffluence of the stars in the stream. Once in the stream the stars are no longer bound to the cluster, their intrinsic dispersion causes the stream to grow along the orbital direction. Escaping stars also have an intrinsic dispersion, related to the dispersion in the outer cluster region.
Another way of understanding the expansion is by considering the velocity difference between the substructure and the outflowing stars, which depends on the tidal radius. Since the tidal radius is shrinking with time, stars that leave the cluster at later times are slower than stars that left before and this leads to the expansion of the stream. In reality the situation is even more complicated. The stream length is actually oscillating during one orbit, being stretched at pericenter and compressed at apocenter. The linear expansion only acts on average over several orbital periods. Therefore for short timescales, the periodic oscillating effect must be taken into account.

This paper is structured as follows: In section 2 we construct a simplified model for a tidal stream and we find a relation between the stream density and its velocity dispersion. Sections 3 and 4 are dedicated to the study of the stream stability, where we first derive the linearised equation of perturbations and then look at the one dimensional collapse along the stream direction. In section 5 we take a critical look at our model by comparing with the detailed dynamics of streams using N-body simulations. The orbital oscillation of the stream length and its influence on collapse are considered. Finally we give our conclusions in section 6.

\section{Modelling a tidal stream}
The general case of a streaming cluster is a problem of many particle dynamics that can be solved self-consistently with simulations. Analytical statements can be made by considering a model with simplifying assumptions. Thereby one has to be careful to avoid over-simplification. We model a tidal stream as a self gravitating cylinder of collisionless matter with an expansion in the direction of the cylinder axis. For the cluster as well as for the host we choose isothermal spheres so that we can use the simplifying relations
\begin{equation}\label{isorelations}
\frac{r_t}{R}\sim c\left(\frac{m}{M}\right)^{1/3}\sim c^{3/2}\left(\frac{\sigma_{cl}}{\sigma_{gal}}\right),\hspace{0.5cm}GM=2\sigma_{gal}^2 R,
\end{equation}
where $m$, $M$ and $\sigma_{cl}$, $\sigma_{gal}$ are the masses respectively the velocity dispersions of the cluster and the host. The distances $r_t$ and $R$ are the tidal radius of the cluster and the orbital radius to the host. For an isothermal sphere the correction factor $c\sim0.8$ \citep{Binney2008}.

Whilst many systems can be reasonably well described by an isothermal potential over the scales of interest, our results would not apply to systems orbiting within very different potentials. For example, a star cluster within a constant density potential would not even produce streams. However, for Palomar 5 and for many of the streams in our Galactic halo, or within galaxy clusters, an isothermal potential is a good approximation over the range $0.01-0.5R_{virial}$ \citep{Klypin2002}.

\begin{figure}
\centering
\includegraphics[scale=0.85]{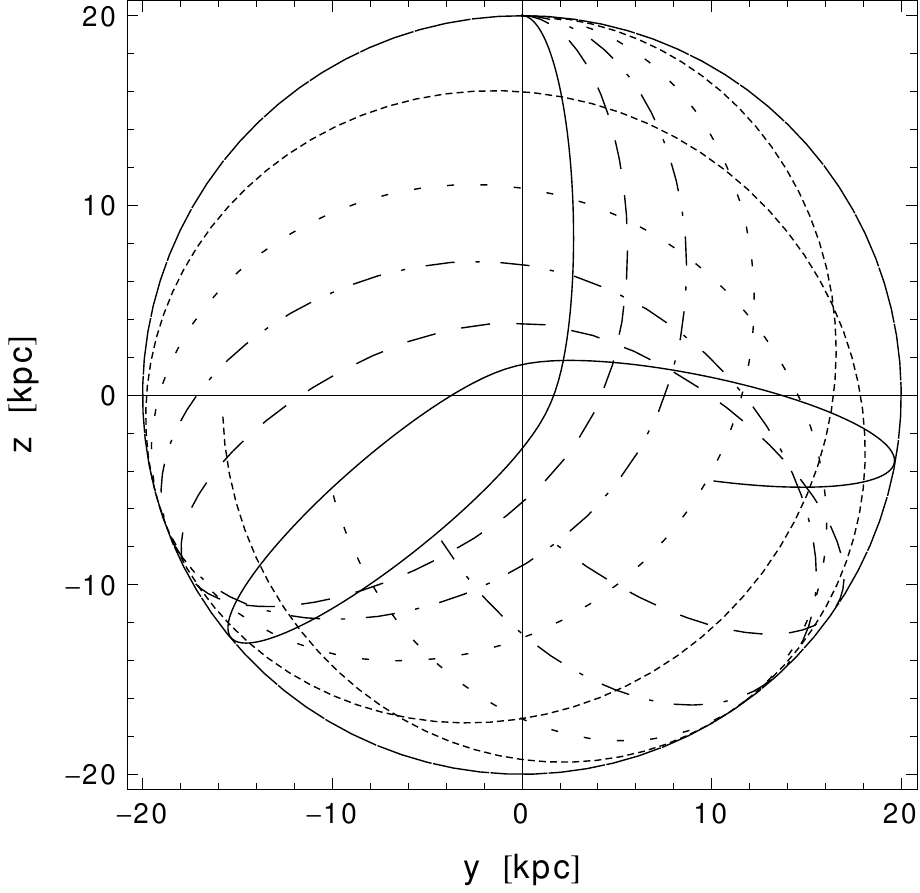}
\caption{Orbits of the star clusters in our simulations. The eccentricity is given in terms of the parameter $b$ defined as $\dot{R}=b V$ (with $b=0$ for a circular orbit and $b=1$ for a radial infall). In increasing eccentricity: continuous ($b=0$), narrow-dotted ($b=0.14$) broad-dotted ($b=0.34$), dashed-dotted ($b=0.54$) dashed ($b=0.74$) and continuous ($b=0.88$).}
\label{orbits}
\end{figure}

In order to test the basic assumptions of our model we perform simulations of a star cluster orbiting with different eccentricities within an isothermal host potential. The velocity dispersion is chosen to be $\sigma_{cl}=4$ km/s for the cluster and $\sigma_{gal}=200$ km/s for the host. Every simulation starts with the star cluster at a radius of 20 kpc and we choose different perpendicular initial velocities from 283 km/s for a circular orbit to 50 km/s for the most eccentric orbit. The global potential is a fixed analytic potential whilst the star cluster is modelled using $2\times10^5$ stars, set up in an equilibrium configuration at the starting position \citep{Zemp2008}. The evolution is followed using the N-body code PKDGRAV \citep{Stadel2001}, adopting high precision parameters for the force accuracy. The softening length of the star particles is $\epsilon=0.005$ kpc. All simulated orbits are illustrated in Fig. \ref{orbits}.

There are two mechanism responsible for the stream growth, on the one hand, the outflow of matter leaving the cluster with a certain velocity difference $\Delta V$ and on the other hand the stream expansion due to diffluence of the initial dispersion. The expansion velocity $w$ is given by $w\sim 2\sigma_{cl}$, which corresponds to the diffluence velocity of a bunch of particles leaving the cluster at the same time. Using conservation of angular momentum $L$ leads to the velocity difference $\Delta V$:
\begin{equation}\label{deltav}
L=RV\sin\theta=(R+r_t)(V-\Delta V)\sin\theta\hspace{0.2cm}\Rightarrow\hspace{0.2cm}\frac{\Delta V}{V}\sim \frac{r_t}{R}
\end{equation}
Here we have assumed $R\gg r_t$ and $V\gg\Delta V$. In an isothermal potential the value of $V$ must be somewhere between $V_R=(4/\pi)^{1/2}\sigma_{gal}$ and $V_c=2^{1/2}\sigma_{gal}$, the radial and circular velocities. Using (\ref{isorelations}) we therefore obtain $\Delta V\sim \sigma_{cl}$ as well as
\begin{equation}\label{velocityrelation}
w\sim 2\Delta V.
\end{equation}

Physically this means that all particles belonging to the stream at $t_0$ will be distributed over the entire stream length at all time $t>t_0$. Or in other words, even if there is no more outflow from the cluster, the stream always stays attached to the cluster.

\begin{figure}
\centering
\begin{minipage}{8cm}
\includegraphics[scale=0.58]{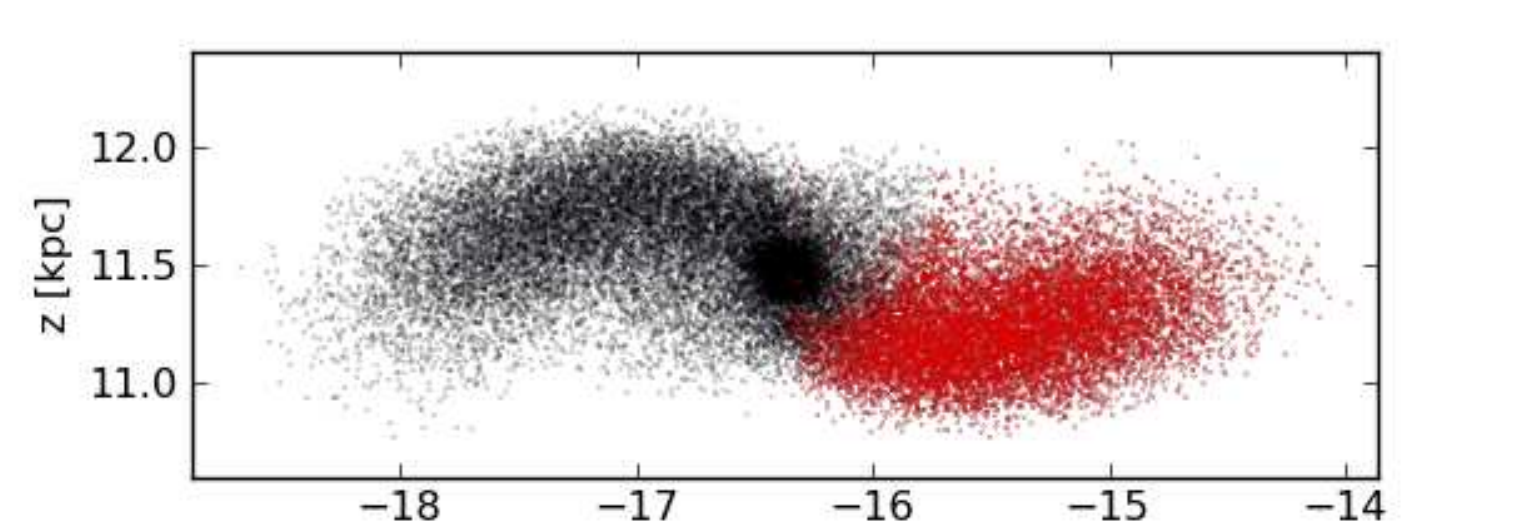}
\end{minipage}
\begin{minipage}{8cm}
\includegraphics[scale=0.63]{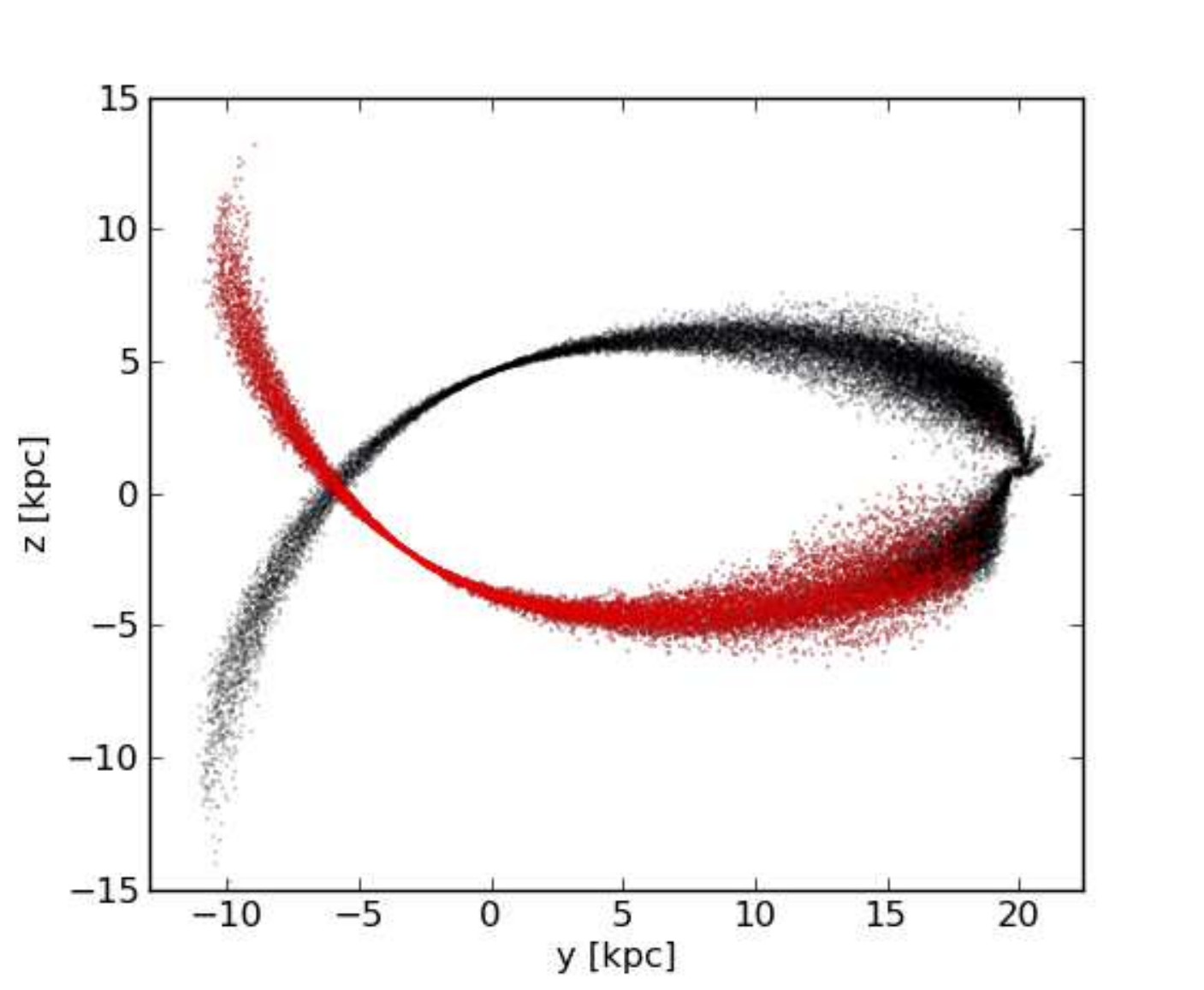}
\end{minipage}
\caption{\textit{Simulation of an isothermal cluster with an eccentricity of $b = 0.74$. The image on the top shows the cluster after one orbit where the particles of one stream are marked in red. The image on the bottom shows the cluster after nine orbits with the distribution of the particles marked before.}}
\label{marked_particles}
\end{figure}

The amount of diffluence can be estimated in the simulation by marking particles at a certain time $t_0$ and looking at where they are found in the stream at $t\gg t_0$. The first image of Fig. \ref{marked_particles} shows a cluster at apocenter after one orbit (195 Myr) with the particles of one stream marked in red. In the second image we see the cluster at apocenter after nine orbits (1756 Myr) along with the distribution of the particles marked before. The particles marked at the early time are located throughout the stream at later times, confirming our above statement.

The width of the stream depends on the velocity dispersion $\sigma$. A particle with an energy excess during the outflow will be on an orbit with a slightly different eccentricity and will therefore complete a full oscillation within the stream during one orbital time $T$. The radius corresponding to half of the stream width is then approximately given by
\begin{equation}\label{streamwidth}
r_{\bot}\sim\frac{1}{2}\sigma T.
\end{equation}
On the other hand the length of the stream after one orbit is simply
\begin{equation}\label{length}
l_0\sim wT\sim 2\sigma T,
\end{equation}
assuming the approximate relation $\sigma _{cl}\sim\sigma$. After one orbit a single stream should therefore be about twice as long as it is wide. This is the case in all our simulations and can be checked in the first image of Fig \ref{marked_particles}.

The linear density of a stream is given by the relation
\begin{equation}\label{mu}
\mu=\frac{dm}{dz}=\frac{\dot{m}}{\dot{z}}\sim\frac{\dot{m}}{2\Delta V},
\end{equation}
Here we have used $\dot{z}\sim w\sim 2\Delta V$, what results in an additional factor of two compared to a static stream because of the stretching effect of the expansion. The rate of outstreaming matter is estimated to be
\begin{equation}\label{dotm}
\dot{m}=\frac{(m_a-m_p)}{T}\sim\frac{2c^{\frac{3}{2}}}{T}\frac{\sigma_{cl}^3}{G\sigma_{gal}}(R_a-R_p)\sim\frac{c^{\frac{3}{2}}}{G}\frac{\sigma_{cl}^3}{\sigma_{gal}}\dot{R},
\end{equation}
where we have used the relations (\ref{isorelations}). The outflow of the matter is averaged over one orbital period. With the relation $\dot{R}=bV$, where the parameter $b$ depends on the cluster orbit (with $b=0$ for a circular orbit and $b=1$ for a radial infall), the linear density becomes
\begin{equation}
\mu\sim\frac{c^{3/2}}{2G}\frac{\sigma_{cl}^3}{\sigma_{gal}}\left(\frac{V}{\Delta V}\right)b\sim\frac{\sigma_{cl}^2b}{2G},
\end{equation}
and the Toomre parameter is then given by
\begin{equation}\label{Toomreparameter}
q\equiv\frac{\sigma^2}{2G\mu}\sim\frac{1}{b}.
\end{equation}
The smallest value for the Toomre parameter is therefore $q\sim1$ which corresponds to a radial orbit. The Toomre parameter of relation (\ref{Toomreparameter}) is four times larger than the one obtained by \citet{Quillen2010}, the reason being a factor of two which comes in at equation (\ref{mu}) as well as the averaging of the mass outflow in equation (\ref{dotm}). Both effects are directly related to the expansion of the stream, not considered by Quillen and Comparetta.

An independent way to calculate the Toomre parameter is by using the virial theorem for an isothermal sphere, truncated at the tidal radius $r_t$:
\begin{equation}
\sigma_{cl}^2=\frac{|W|}{m_{cl}}=\frac{4\pi G}{m_{cl}}\int_{0}^{r_t}dr r \rho(r)M(r)=\frac{4\sigma_{cl}^4 r_t}{Gm_{cl}},
\end{equation}
\begin{equation}
\sigma_{cl}^2=\frac{Gm_{cl}}{4 r_t}.
\end{equation}
Using the approximation $\sigma_{cl}\sim\sigma$ then leads to
\begin{equation}
q=\frac{\sigma^2l_0}{2Gm_{st}}\sim\frac{1}{8}\frac{m_{cl}}{m_{st}}\frac{l_0}{r_{t}}.
\end{equation}
For the extreme case of a radial orbit $m_{cl}\sim 2 m_{st}$ and we obtain $q\sim 1$. This means that for all orbits $q$ must be larger than one, a result that confirms the relation (\ref{Toomreparameter}) above.

The Toomre parameter can also be determined in the simulations by measuring the velocity dispersion and the linear density. However, it turns out that the dispersion is very difficult to quantify accurately because over one orbital period it strongly fluctuates at any Lagrangian point (for example, around any star). This is due to the oscillation of stream-length and stream width, which happens because the particles in the stream are on nearly free orbits around the host. The velocity dispersion therefore is affected by the number of stars used in its measurement since that changes the region of the stream over which the dispersion is calculated.

\begin{figure}
\centering
\includegraphics[scale=0.8]{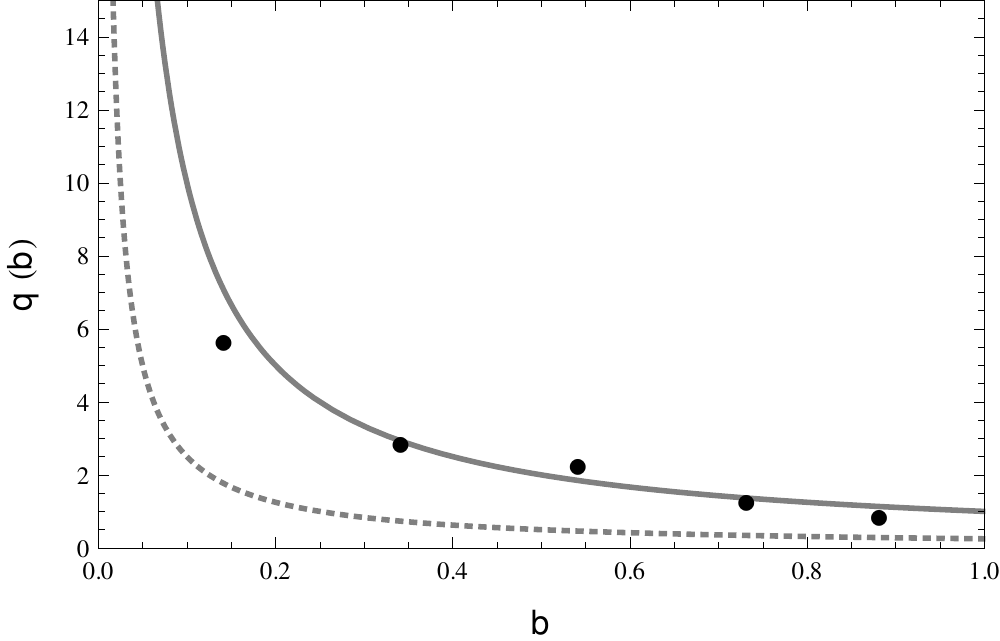}
\caption{The Toomre parameter as a function of the eccentricity parameter $b$. The black dots are the measurements from the different simulations. The solid gray line corresponds to equation (\ref{Toomreparameter}), while the dotted line is the prediction from \citet{Quillen2010}.}
\label{q(b)}
\end{figure}

In Fig. \ref{q(b)} we plot the Toomre parameter, where the dispersion is measured in the middle of the stream, at apocenter after one orbital period and assuming an isotropic distribution (looking at the ration of tangential to radial velocity dispersions  after one orbit we can see that this assumption is approximately valid). We notice that the simulations roughly follow the theoretical prediction which is given by the solid gray line.

Already at that stage of our analysis it becomes clear that the expansion has a strong stabilising effect because it leads to a significant boost of the Toomre parameter. In the next section we will see that the stability of a stream is additionally enforced, since the expanding environment leads to a damping in the the evolution of perturbations.

\section{Perturbations in an expanding cylinder}
In order to find a criterion for the stability, we are now modeling a tidal stream as a non-rotating elongated cylinder of collisionless matter that is linearly expanding in the direction of its long axis. For the expansion we introduce the comoving coordinate $s=az$ with $a(t)=\alpha t$ and set $z=l_0$, where $l_0$ is the stream length after one orbital period $T$. The expansion factor then becomes
\begin{equation}\label{alpha}
\alpha=\frac{1}{T}.
\end{equation}
The orbital period is a natural time measure since the outstreaming from the cluster into the tails is mainly happening during the cluster orbit from apo- to pericenter when the tidal radius is shrinking. During the other half of the orbit the tidal radius is growing again and there is nearly no streaming mass loss.

An analytical treatment of the stability of an expanding cylinder is possible either on scales much smaller or much larger than the cylindrical radius. In the former case we can treat the fluid as homogeneous and we therefore get the usual Jeans length
\begin{equation}\label{homJeanslength}
\lambda^h_J=\sqrt{\frac{\pi \sigma^2}{G\rho}}.
\end{equation}
With the relation (\ref{Toomreparameter}) as well as the linear density $\mu=\pi r_{\bot}^2\rho$ we then obtain
\begin{equation}
\frac{\lambda^h_J}{r_{\bot}}=\sqrt{2\pi^2 q}\sim\sqrt{\frac{2\pi^2}{b}}.
\end{equation}
Since the eccentricity parameter $b$ is always larger than one, the Jeans length exceeds the radius of the cylinder and we can exclude collapse on scales smaller than $r_{\bot}$.

However there is still the possibility of collapse in the longitudinal direction of the cylinder on scales larger than $r_{\bot}$. This is the second analytically treatable case which leads to a very different stability criterion. In order to determine the behaviour of longitudinal perturbations we are now going to derive the equations for the evolution of density perturbations. This is usually done by integrating and linearising the collisionless Boltzmann equation \citep{Peebles1980}. Since we are looking at a thin cylinder, we can assume a phase-space density of the form
\begin{equation}\label{phasespacedensity}
f(z,p,t) = \left\lbrace \begin{array}{cc} a[\rho_b+\rho_1(z,t)]f(p), & r<r_{\bot} \\ 0, & r>r_{\bot} \end{array} \right.
\end{equation}
Here we have introduced a homogeneous background density $\rho_b$ as well as a first order perturbation $\rho_1$. An integration of the phase-space density immediately leads to the stream density
\begin{equation}
\rho= \frac{1}{a}\int dpf(z,p,t)=\frac{(\mu_b+\mu_1)}{\pi r_{\bot}^2}=\frac{1}{\pi r_{\bot}^2}\frac{\mu_0}{a}(1+D),
\end{equation}
where $D=\mu_1/\mu_b$ is the dimensionless overdensity.

The evolution of the phase-space density is described by the one dimensional collisionless Boltzmann equation with expanding coordinate
\begin{equation}\label{Boltzmann}
\frac{\partial}{\partial t}f(z,p,t) + \frac{p}{a^2}\frac{\partial}{\partial z} f(z,p,t) - \frac{\partial\Phi}{\partial z}\frac{\partial}{\partial p}f(z,p,t)=0.
\end{equation}
It is now straightforward to derive the continuity and the momentum equation of the stars in the cylinder. They are given by
\begin{equation}
\partial_t(1+D) + \frac{1}{a}\partial_z\left[\langle v\rangle(1+D)\right] = 0,
\end{equation}
\begin{equation}
\partial_t\left[a\left<v\right>(1+D)\right] + \partial_{z}\Phi(1+D) + \partial_z\left[\langle v^2\rangle(1+D)\right] = 0,
\end{equation}
where
\begin{equation}\label{avvelavdisp}
\langle v\rangle=\frac{\int p f dp}{a\int fdp},\hspace{1cm}\langle v^2\rangle=\frac{\int p^2fdp}{a^2\int fdp}.
\end{equation}
By substituting the derivative of the second equation into the first we finally find the equation of perturbation:
\begin{equation}\label{deltaequation}
\partial_t^2 D + 2\frac{\dot{a}}{a}\partial_t D = \frac{1}{a^2}\partial_z\left[(1+D)\partial_z\Phi\right] + \frac{1}{a^2}\partial_z^2\left[(1+D)\langle v^2\rangle\right].
\end{equation}

In order to solve this differential equation we still need to know the potential of a cylinder. The simplest assumption is to take
\begin{equation}
\Phi(r,z,t)=\Phi^{(0)}(r)+\Phi^{(1)}(r,z,t),
\end{equation}
\[
\Phi^{(1)}(r,z,t)=\phi^{(1)}(r,t)e^{ik_0z}
\]
\citep{Fridman1984}, where $k_0$ is the comoving wave number in z-direction. The Poisson equation for the zero-order term is simply
\begin{equation}
\frac{1}{r}\frac{d}{dr}\left(r\frac{d\Phi^{(0)}}{dr}\right)=4\pi G \left\lbrace\begin{array}{cc}\rho_0, & r<r_{\bot} \\ 0, & r>r_{\bot}\end{array}\right.
\end{equation}
with the solution $\Phi^{(0)}(r)=\pi G \rho_0 r^2 + const$. In contrast to a self gravitating cylinder, a stream is embedded in the dominating potential of the host and the zero order term looks different. However, a dependence of the potential in the z-direction only comes in as a first order effect due to the internal structure of the stream. Therefore we obtain the following Poisson equation at first order
\begin{equation}
\partial_r^2\Phi^{(1)} + \frac{1}{r}\partial_r\Phi^{(1)} - \frac{k_0^2}{a^2}\Phi^{(1)} = \left\lbrace\begin{array}{cc}4\pi G \rho_1, & r<r_{\bot} \\ 0, & r>r_{\bot}\end{array}\right.
\end{equation}
where $\rho_1$ may vary along the axis of the cylinder. Two independent solutions of this homogeneous differential equation are the modified Bessel equations of first and second kind $I_0[x]$ and $K_0[x]$. The general inner and outer solution are given by
\begin{equation}
\Phi_{<}^{(1)}(r) = AI_0\left[\frac{k_0 r}{a}\right]+ BK_0\left[\frac{k_0 r}{a}\right] - \frac{4\pi G \rho_1 a^2}{k_0^2},
\end{equation}
\begin{equation}
\Phi_>^{(1)}(r) = A'I_0\left[\frac{k_0 r}{a}\right]+ B'K_0\left[\frac{k_0 r}{a}\right],
\end{equation}
where the boundary conditions require $B=A'=0$. With the matching conditions $\Phi_<^{(1)}(r_{\bot})=\Phi_>^{(1)}(r_{\bot})$ and $\partial_r\Phi_<^{(1)}(r_{\bot})=\partial_r\Phi_>^{(1)}(r_{\bot})$ we find
\begin{equation}
A=\frac{4\pi G \rho_1 a^2 K_0'\left[k_0r_{\bot}/a\right]}{k_0^2W\left[k_0 r_{\bot}/a\right]},\hspace{0.2cm}W = -\frac{k_0}{a}(I_0K_1 + I_1K_0).
\end{equation}
We now look at the case of large perturbations in a thin stream ($k_0r_{\bot}/a<<1$). In the asymptotic limit we get
\begin{equation}
A \simeq \frac{2G\mu_0}{a}\left[\frac{2a^2}{(k_0r_{\bot})^2} + \gamma + \log\left(\frac{k_0r_{\bot}}{2a}\right)\right]D.
\end{equation}
The first order potential inside the stream is then given by
\begin{equation}\label{potential}
\Phi_{<}^{(1)} \simeq \frac{\sigma_0^2}{q_0 a}\left[\log\left(\frac{k_0r_{\bot}}{2a}\right)+\gamma\right]D,
\end{equation}
where we have used $\mu=\pi r_{\bot}^2\rho$ together with relation (\ref{Toomreparameter}). The Euler constant is $\gamma=0.577$.

Using (\ref{deltaequation}) and (\ref{potential}) we obtain a closed set of equations for the perturbations $D$ that can now be linearised. We therefore set $D<<1$, as well as $\langle v^2\rangle(z,t) = \sigma^2(t) + O(v_1^2)$ which gives
\begin{equation}
\ddot{D} + 2\frac{\dot{a}}{a}\dot{D} = \frac{1}{a^2}\partial_z^2\Phi_<^{(1)} -\frac{k_0^2\sigma^2}{a^2}D,
\end{equation}
\begin{equation}\label{lindeltaequation0}
\ddot{D} + 2\frac{\dot{a}}{a}\dot{D} = -\frac{k_0^2}{a^2}\left\lbrace \frac{2G\mu_0}{a}\left[\log\left(\frac{k_0r_{\bot}}{2a}\right)+\gamma\right] + \sigma^2\right\rbrace D.
\end{equation}

In a collisionless cylinder the longitudinal velocity dispersion decreases as $\sigma=\sigma_0a^{-1}$ (see eq. \ref{avvelavdisp}), while the perpendicular velocity dispersion stays constant. Equation (\ref{lindeltaequation0}) can therefore be written as
\begin{equation}\label{lindeltaequation}
\ddot{D} + 2\frac{\dot{a}}{a}\dot{D} = -\frac{\sigma_0^2 k_0^2}{q_0 a^3}\left\lbrace\log\left(\frac{k_0r_{\bot}}{2a}\right) + \gamma + \frac{q_0}{a}\right\rbrace D.
\end{equation}
The perturbation, $D$, is damped if the right hand side of equation (\ref{lindeltaequation}) is negative. Therefore we can define a Jeans length
\begin{equation}\label{Jeanslength}
\lambda_J=\pi r_{\bot} \exp\left(\frac{q_0}{a}+\gamma\right),
\end{equation}
which is very different from the stability criterion in a homogeneous surrounding (\ref{homJeanslength}). The geometry of a thin cylinder leads to a Jeans length with an exponential form that guarantees stability up to much larger scales.

\begin{figure}
\centering
\begin{minipage}{4cm}
\includegraphics[scale=0.4]{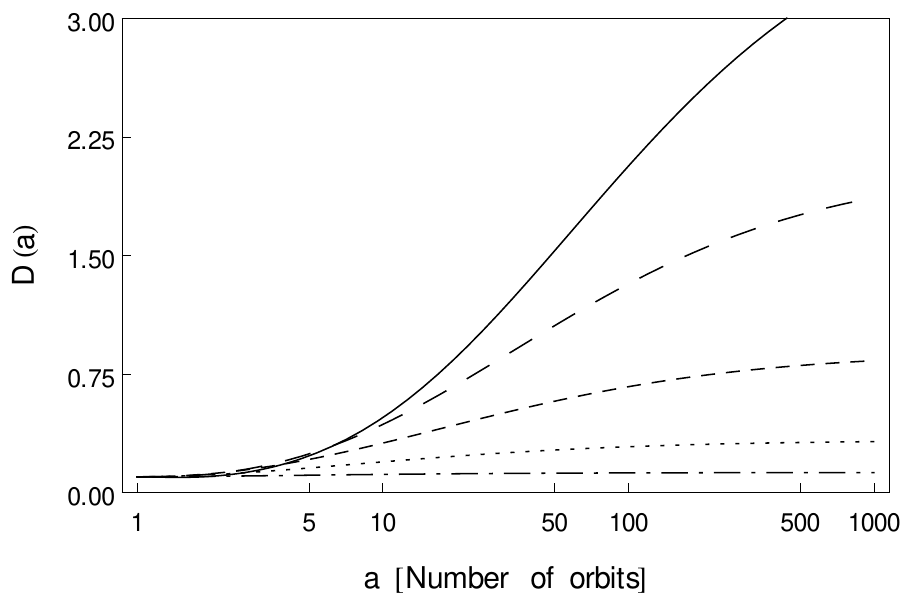}
\end{minipage}
\begin{minipage}{4cm}
\includegraphics[scale=0.4]{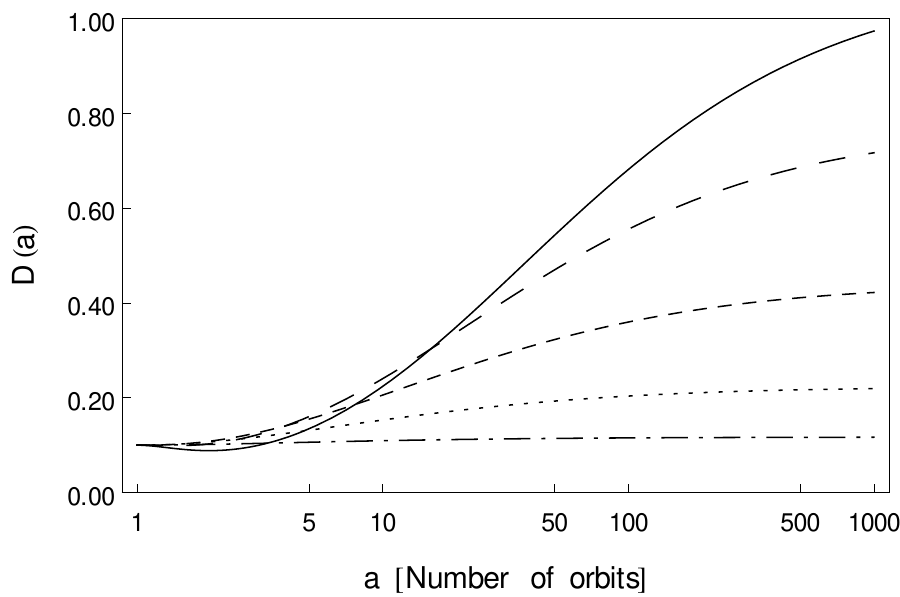}
\end{minipage}
\begin{minipage}{4cm}
\includegraphics[scale=0.4]{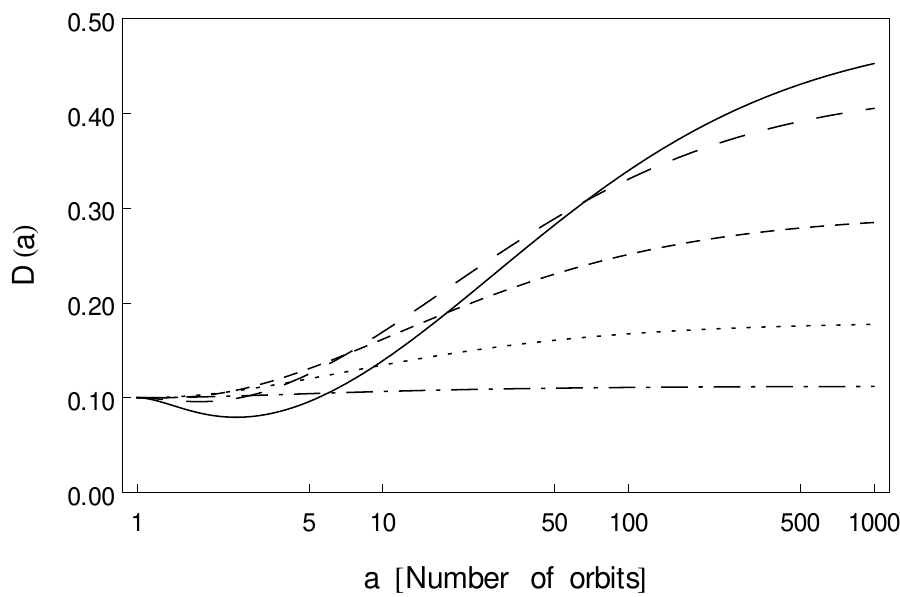}
\end{minipage}
\begin{minipage}{4cm}
\includegraphics[scale=0.4]{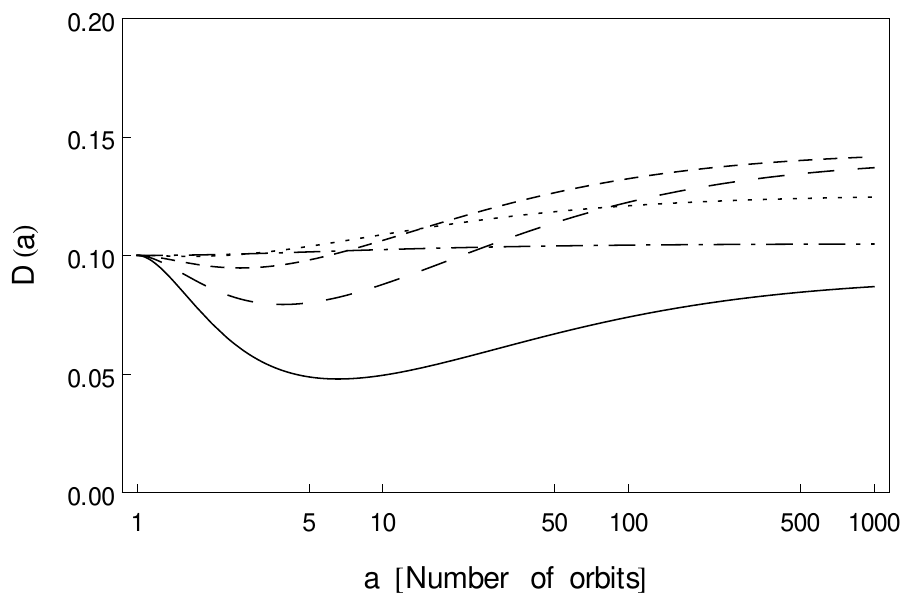}
\end{minipage}
\caption{\textit{Dynamics of the perturbation $D$ with respect to the scale factor $a$ for the comoving factors $k_0r_{\bot}=$ 0.9 (solid), 0.7 (wide-dashed), 0.5 (narrow-dashed) 0.3 (dotted) and 0.1 (dashed-dotted). From top left to bottom right: $q=$ 0.5, 0.75, 1, 2}}
\label{perturbations}
\end{figure}

Equation (\ref{lindeltaequation}) can now be simplified using (\ref{streamwidth}) and taking $a$ as variable:
\begin{equation}\label{lindeltaequation2}
D''(a) + \frac{2}{a}D'(a)  =
\end{equation}
\[
-\frac{4 (k_0r_{\bot})^2}{q_0 a^3}\left\lbrace\log\left(\frac{k_0r_{\bot}}{2a}\right) + \gamma + \frac{q_0}{a}\right\rbrace D(a),
\]
There are two remaining free parameters, namely $q_0$ and $k_0r_{\bot}$, which describe the eccentricity of the orbit and the scale of the perturbation compared to the width of the stream. In Fig. \ref{perturbations} we plotted the numerical solutions for different sets of parameters. For a small Toomre parameter the perturbations will become nonlinear and we expect gravitational collapse to occur. However, for larger $q$ the perturbations either undergo a damped oscillation or they freeze out after an unsubstantial phase of growth. Comparing these results with the relation (\ref{Toomreparameter}) leads to the conclusion that the Toomre parameter of a tidal stream is always large enough to assure stability in all cases of interest.
\begin{figure}
\centering
\begin{minipage}{4cm}
\includegraphics[scale=0.4]{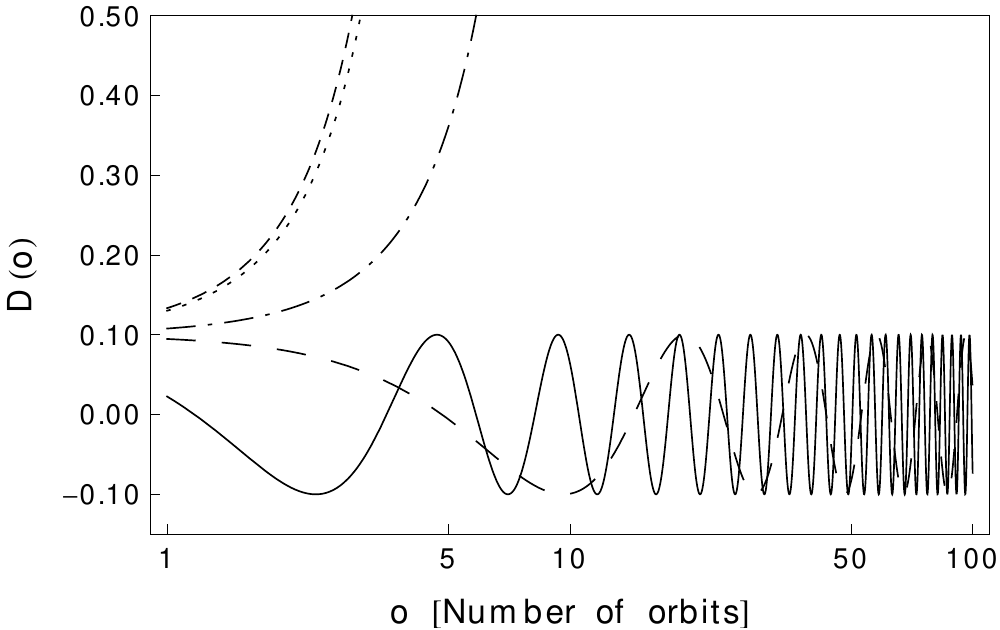}
\end{minipage}
\begin{minipage}{4cm}
\includegraphics[scale=0.4]{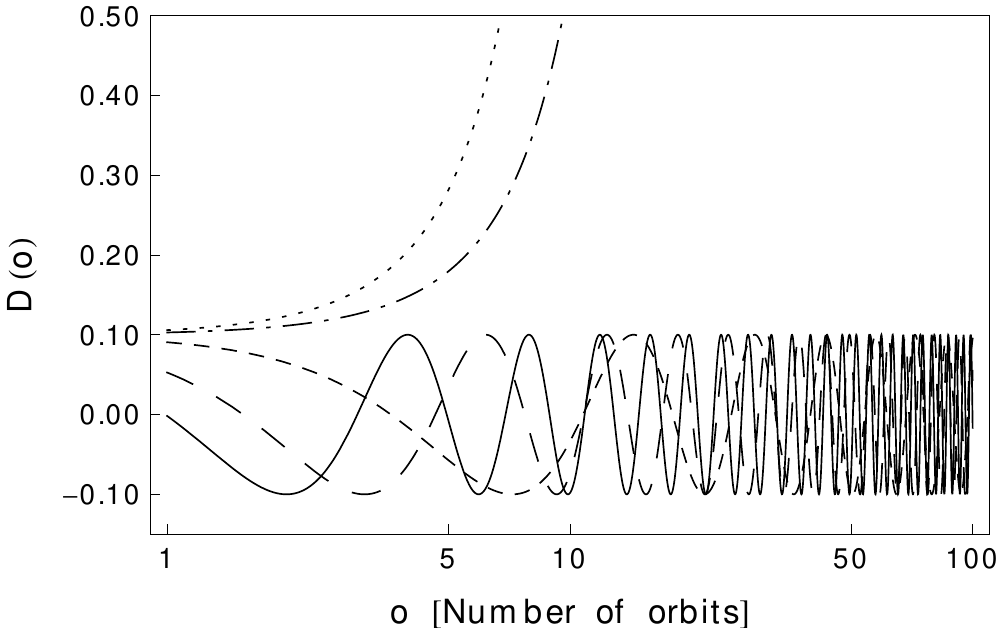}
\end{minipage}
\begin{minipage}{4cm}
\includegraphics[scale=0.4]{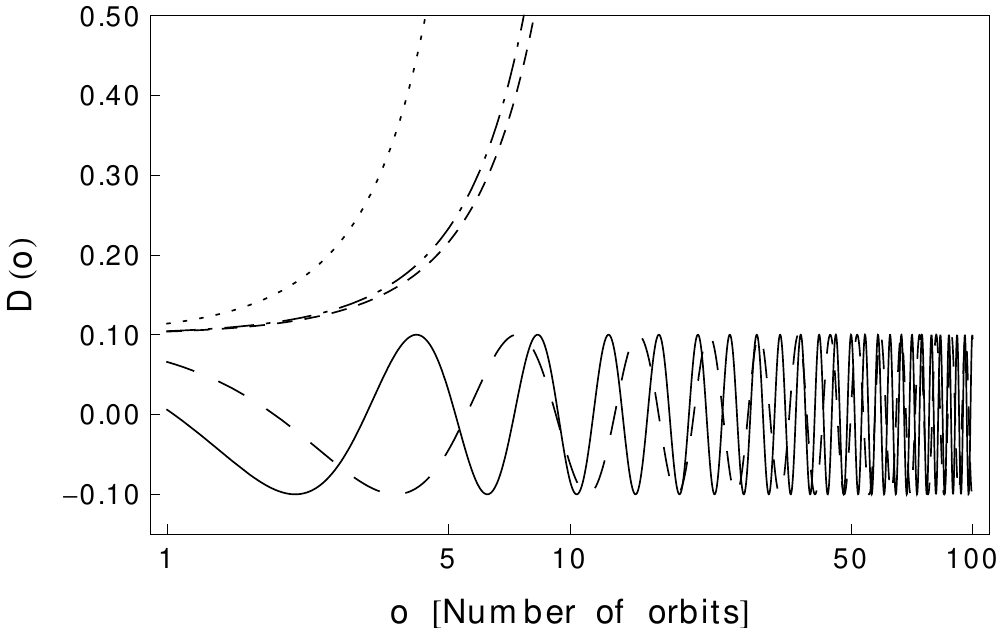}
\end{minipage}
\begin{minipage}{4cm}
\includegraphics[scale=0.4]{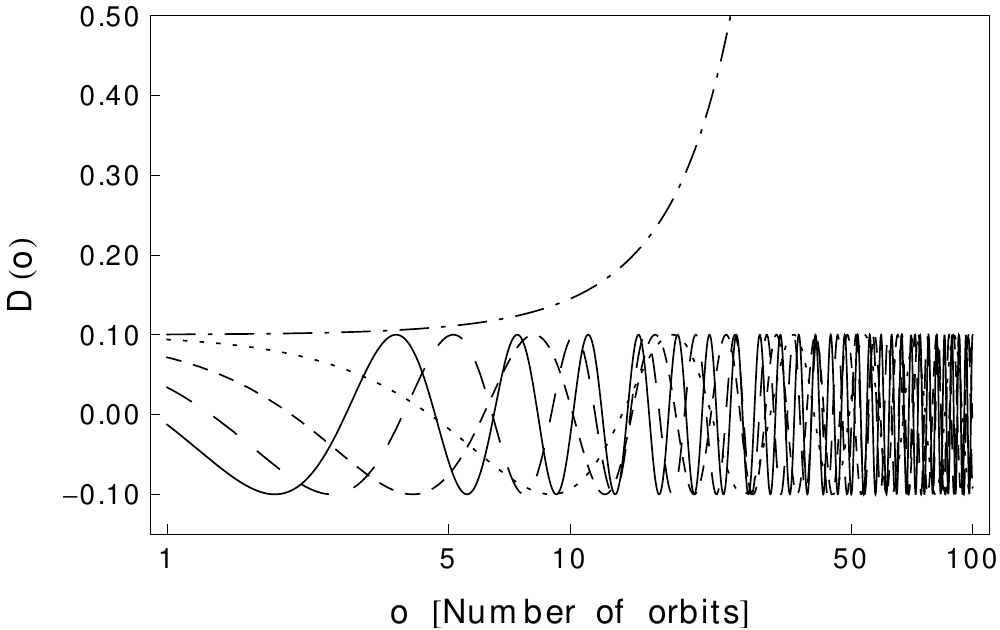}
\end{minipage}
\caption{\textit{Perturbations $D$ in the case of a static cylinder, where $o$ is the number of orbits. The different lines represent the factors $kr_{\bot}=$ 0.9 (solid), 0.7 (wide-dashed), 0.5 (narrow-dashed) 0.3 (dotted) and 0.1 (dashed-dotted) in static coordinates. From top left to bottom right: $q=$ 0.5, 0.75, 1, 2}}
\label{staticperturbations}
\end{figure}

In order to see the effect due to the linear expansion, we also look at the case of a static cylinder. The evolution of perturbations is then given by equation (\ref{lindeltaequation}) with $a=1$ and $\dot{a}=0$ and its behaviour is plotted in Fig. \ref{staticperturbations}.
Even for a large Toomre parameter $q$, there are always collapsing modes supposing an infinitely extended cylinder. This is fundamentally different in the expanding case, where all modes are damped for a high enough $q$, leading to stability on all scales.

Until now we analysed the stability of a stream with linear perturbation theory. In the next section we take a different look at the stream stability by exploring the longitudinal collapse of cylindrical slices. This somehow more heuristic approach is not restricted to the linear regime and gives an independent analysis of the problem.

\section{Shell collapse in a cylinder}
In a one dimensional case of an extended cylinder the spherical collapse reduces to the longitudinal collapse of thin slices. We therefore consider a homogeneous and infinitely long expanding cylinder with a top hat perturbation at the time $t_i$. The stream can then be cut into slices, which evolve at constant energy. The energy at a certain distance $s$ is given by
\begin{equation}
E_i= \frac{1}{2}v_i^2 + \Phi(s)=\frac{1}{2}\left(\frac{\dot{a}_i}{a_i}\right)^2s^2 -\frac{GM_i(s)}{s}.
\end{equation}
Since the mass $M_i$ evolves as
\begin{equation}
M_i(s) = \int (1 + \delta)\mu_b(t_i) ds = \frac{2\mu_0}{a_i}(1 + \delta)s
\end{equation}
we obtain the energy
\begin{equation}\label{energy}
E_i = \frac{1}{2}\alpha^2\left(\frac{s}{a_i}\right)^2 - 2G\frac{\mu_0}{a_i}(1+\delta).
\end{equation}
Slices with a positive total energy will never collapse and therefore $E_i\geq0$ is our stability condition. Equation (\ref{energy}) then leads to
\begin{equation}\label{stabcon0}
s \geq r_{\bot}\sqrt{\frac{8(1+\delta)a_i}{q_0}},
\end{equation}
where we have used the definition of the Toomre parameter (\ref{Toomreparameter}). Slices further away are stable while nearby ones will collapse. The critical distance below which the stream becomes unstable is growing with the square root of time.

In the picture of shell collapse the velocity dispersion is completely ignored, since the diffusion of particles into other slices makes the problem much more complicated. We will however account for the dispersion by an \textit{ad-hoc} introduction of the Jeans length $\lambda_J$, which guaranties the stream stability on small scales. With (\ref{stabcon0}) and (\ref{Jeanslength}) we can then construct the stability criterion
\begin{equation}
q_0 \geq \frac{8(1+\delta)}{\pi^2}a_i e^{-2(q_0a_i^{-1}+\gamma)},
\end{equation}
which is fulfilled at the beginning ($a=1$) and may be violated at some later times ($a>a_c$).
This means that for $t_i=t_0$ all instable slices are below $\lambda_J$ and therefore all the stream is stable. Later on however and depending on $q_0$ unstable modes may appear just above $\lambda_J$.

Since the Jeans length gives a minimum size for the final structure, the initial collapse must start at a scale well above this. A calculation of the collapse-time $t_{coll}$ shows however that $t_{coll}$ dramatically grows with the distance of the slice. Slices only a few times further away than the Jeans length already have a $t_{coll}$ that largely exceeds one Hubble time, at least for $q_0\geq1$. This means that even though there are unstable modes in an expanding stream, they will never have enough time to grow substantially. Collapse only occures for very small values of $q_0$ well below the limit given by (\ref{Toomreparameter}).

This qualitative picture is in agreement with the results plotted in Fig. \ref{perturbations}, where a phase of damped oscillation is followed by a phase of growth, freezing out at a very low level still in the linear regime.

\section{Towards a realistic stream}
A realistic treatment of a tidal stream orbiting its host galaxy can become very complex, which leads us to consider the possibility that our model of an expanding cylinder is an over-simplification and therefore we are missing some important dynamics. In the following we treat possible deviations to our model and discuss their influence on the stability:

\begin{list}{\labelitemi}{\leftmargin=0em}
\item In general, the host galaxy is not simply isothermal, but can have a triaxial shape that varies with time, and it contains substructures. The orbit of a cluster is then no longer within a plane and it may lose its regularity. The analysis of stability effects in such a complex situation is best tackled with full numerical simulations. Nevertheless, there is no a-prior reason to believe that one of these effects could fundamentally alter the stability criterion.

\item A stream approximately traces the orbit of its cluster and is therefore more and more curved the longer it gets. This does not correspond to the straight cylinder used in the model. However the effect of the bending is rather stabilising the stream against longitudinal Jeans instabilities and can therefore confidently be ignored.

\item A much more severe limitation of our model is the assumption of a cylindrical form. In reality streams are often more sheet-like and their thickness strongly varies during the orbital period. The closer a stream approaches the centre of the host, the thinner it gets. The reason for this behaviour is the form of the isothermal host potential which leads to orbits that occupy a narrower real-space volume closer to its centre.
In Fig. \ref{halo_v100} the image of a stream on an eccentric orbit is illustrated. The difference in the stream-width is very pronounced and the sheet like structure at apocenter is also visible. Even though the variation in the thickness has a major influence on the local stream density, it does not affect the longitudinal collapse condition, which only depends on the linear density. Incorporating the effect of the flattening of the stream is somewhat more difficult because it affects the potential (\ref{potential}). However, it is again unlikely that the sheet-like structure would have an enhancing effect on the collapse since it is stretching the stream which reduces its density.
\begin{figure}
\centering
\includegraphics[scale=0.45]{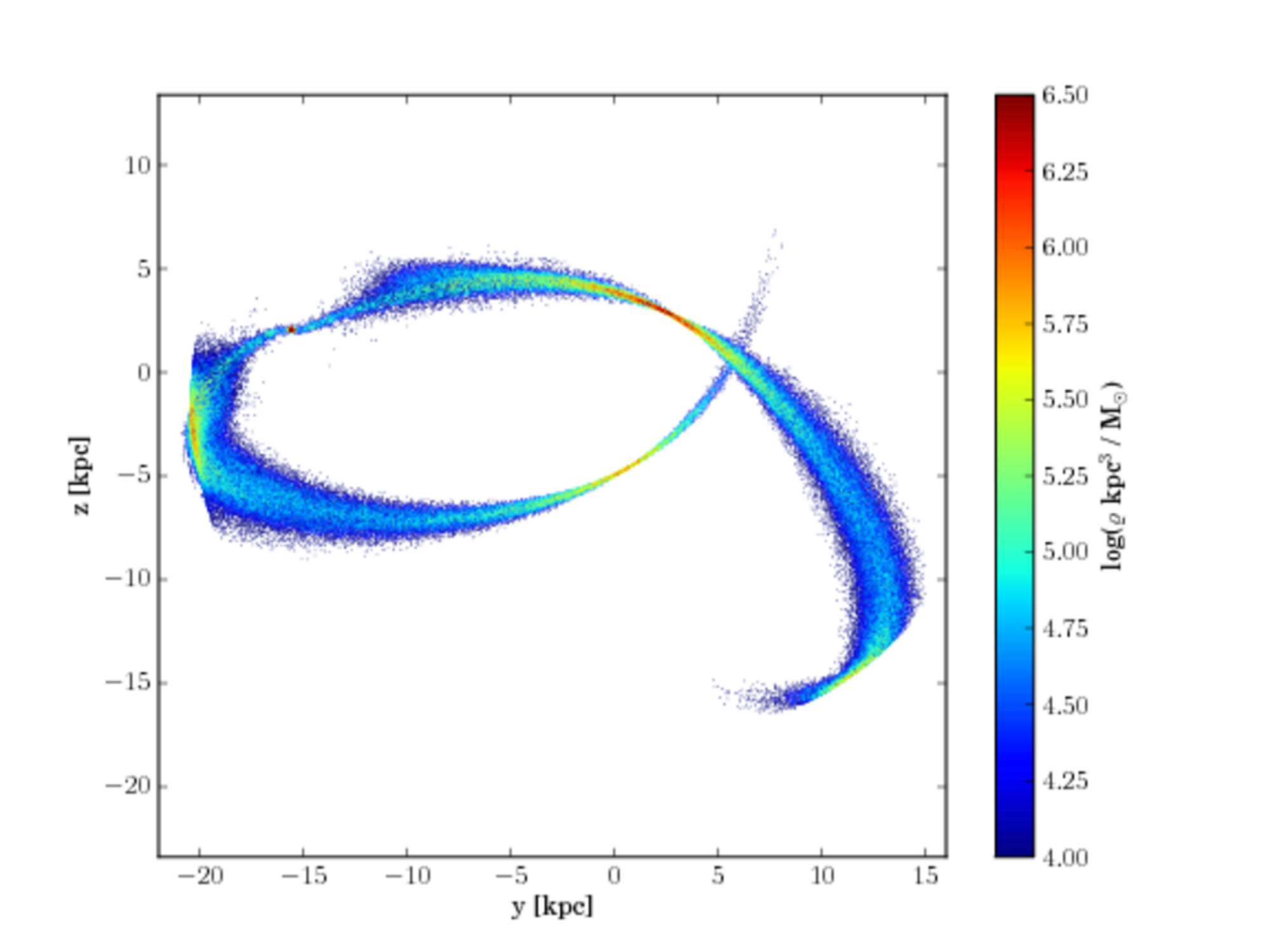}
\caption{Density map of a star cluster with a leading and tailing stream after 2 Gyr in an isothermal host potential. The orbit lies in the (y,z)-plane and has an eccentricity factor of $b=0.74$. The high eccentricity leads to strong variations in the stream width. While the streams are narrower and denseer at pericentre, they become flattend at apocentre with the typical umbrella-like form.}
\label{halo_v100}
\end{figure}

\item As the stream orbits between apocenter and pericenter, its length is oscillating, a fact that is not included in our model assumptions and may affect the stream stability. In fact, the stream only expands linearly on average, its length oscillates during one orbit, being stretched at pericenter and compressed at apocenter. In Fig. \ref{avdis} the average distance of random points in streams on different orbits are illustrated and the orbital oscillation as well as the overall linear expansion are clearly visible.
\begin{figure}
\centering
\includegraphics[scale=0.45]{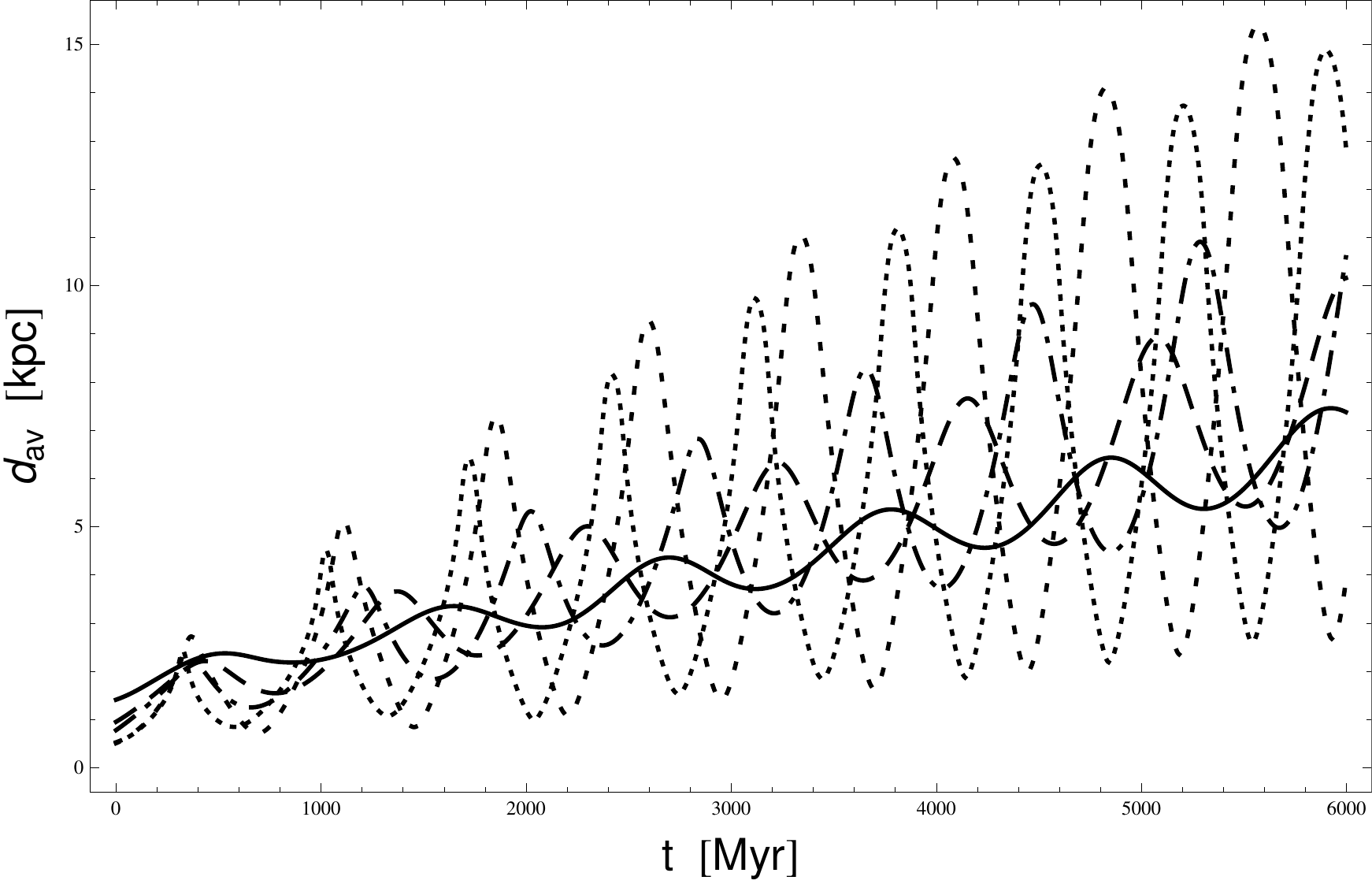}
\caption{The evolution of the distance between chosen particles in the stream for different simulations with $b=0.14$ (full), $b=0.34$ (dashed), $b=0.54$ (dashed-dotted), $b=0.74$ (narrow-dotted) and $b=0.88$ (broad-dotted). Whilst there is linear growth averaged over the orbital motion, the length is oscillating with the orbit, and the amplitude of the oscillation is larger for higher eccentricity. }
\label{avdis}
\end{figure}
These oscillations have an effect on the longitudinal perturbations. From peri- to apocenter, when the stream-length is shrinking, we are no longer in a stable regime and we expect growth. However, this growth happens on a timescale longer than the orbital period so that perturbations do not have time to collapse.
This can be shown by approximating the shrinking of the stream with a linearly decreasing scale factor of the form
\begin{equation}
a(t)=d-\left(d-1\right)\frac{2}{T}t,
\end{equation}
where $d=l_{max}/l_{0}$. The stream length $r(t)=a(t)l_{0}$ now runs from $l_{max}$ to $l_{0}$ in half of an orbital period. We then use the equation of perturbation (\ref{lindeltaequation}) and replace the time variable with the scale factor. The result is
\begin{equation}\label{lindeltaequation3}
D''(a) + \frac{2}{a}D'(a)  =
\end{equation}
\[
-\frac{(k_0r_{\bot})^2}{(d-1)^2q_0 a^3}\left\lbrace\log\left(\frac{k_0r_{\bot}}{2a}\right) + \gamma + \frac{q_0}{a}\right\rbrace D(a),
\]
as well as the initial conditions $D(d)=0.1$ and $D'(d)=0$. We find growing solutions if the right hand side of the above equation is positive, where the actual value determines the growth rate. A large value of $d$ (high eccentricity) gives a small growth factor for a long interval of integration, whilst a small value (low eccentricity) gives a large growth factor for a short interval, the reason being the $d^2$-term in the denominator of (\ref{lindeltaequation3}). Hence, the actual growth of perturbations stays negligibly small in all cases even for a $q$ as low as 0.5 and the overall stability of our streams is therefore ensured. In Fig. \ref{growth} we plotted the evolution of the perturbations between peri- and apocenter for the case of $q=0.5$ and $q=1$ and with $d=2$ and $d=10$.
\begin{figure}
\centering
\begin{minipage}{4cm}
\includegraphics[scale=0.4]{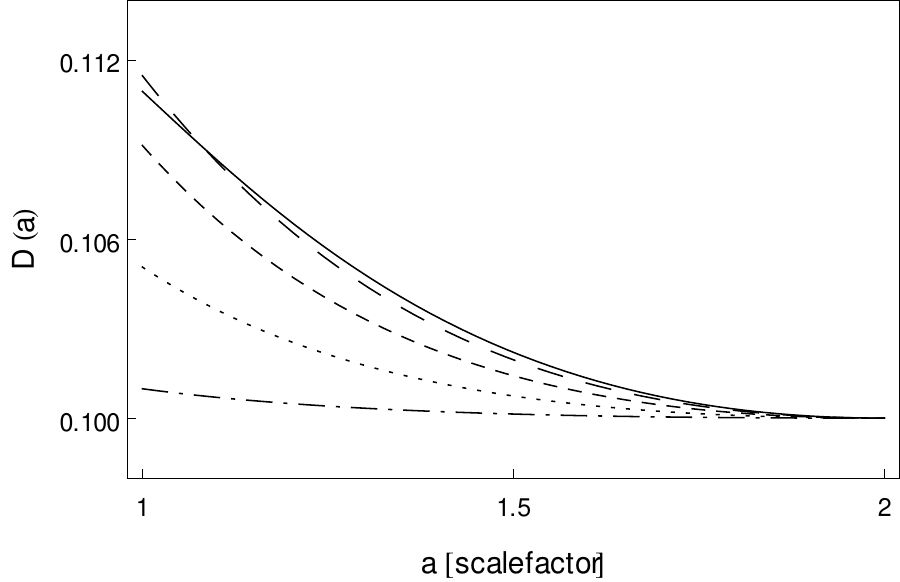}
\end{minipage}
\begin{minipage}{4cm}
\includegraphics[scale=0.4]{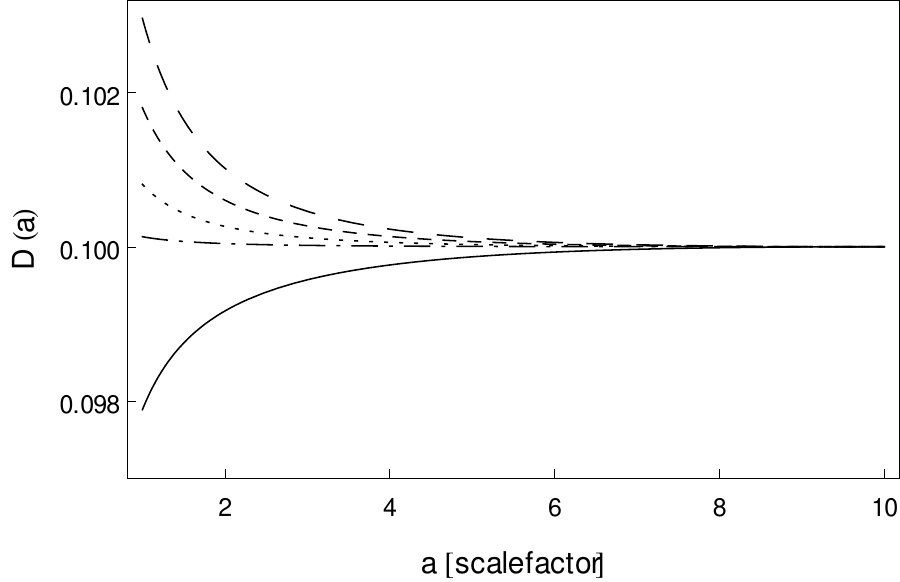}
\end{minipage}
\begin{minipage}{4cm}
\includegraphics[scale=0.4]{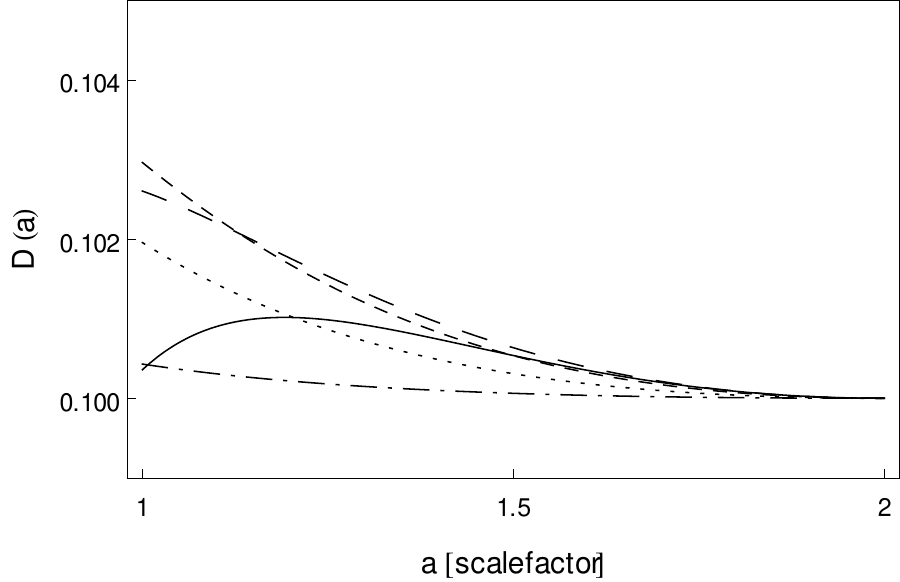}
\end{minipage}
\begin{minipage}{4cm}
\includegraphics[scale=0.4]{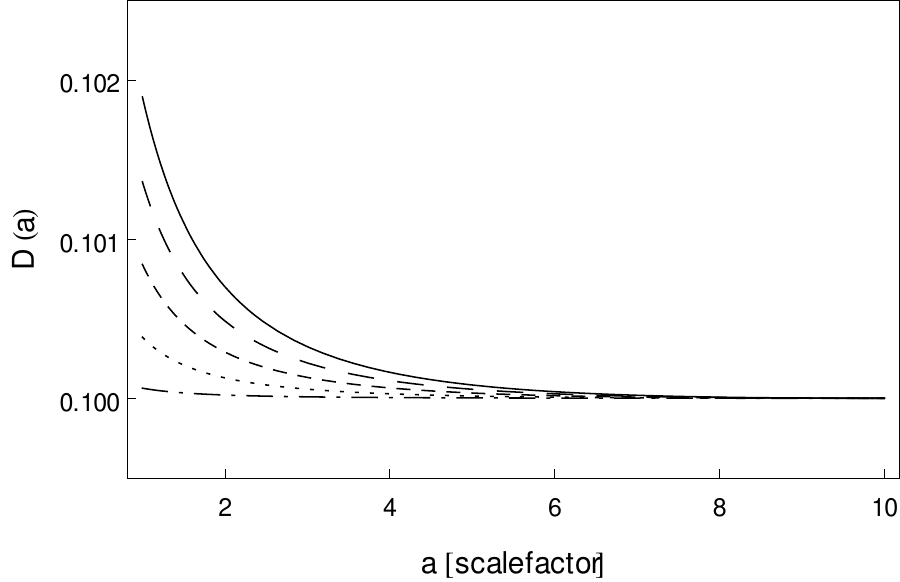}
\end{minipage}
\caption{\textit{Growing perturbations $D$ for a shrinking scale factor $a$ with $k_0r_{\bot}=$ 0.9 (solid), 0.7 (wide-dashed), 0.5 (narrow-dashed), 0.3 (dotted), 0.1 (dashed-dotted). The plots should be read from right to left. Top: $q=0.5$ with $d=2$ (left) and $d=10$ (right). Bottom: $q=1$ with $d=2$ (left) and $d=10$ (right).}}
\label{growth}
\end{figure}
\item Because of the longitudinal contraction at apocenter and the transversal contraction at pericenter the density and the total velocity dispersion are oscillating twice as fast as the stream length. This can be observed in Fig. \ref{sigma}, where we plotted the longitudinal and transversal velocity dispersion of a stream with high orbital eccentricity. The doubling of the frequency comes from the fact that the stream is longitudinally compressed at apocenter and transversely compressed at pericenter.
\begin{figure}
\centering
\includegraphics[scale=0.8]{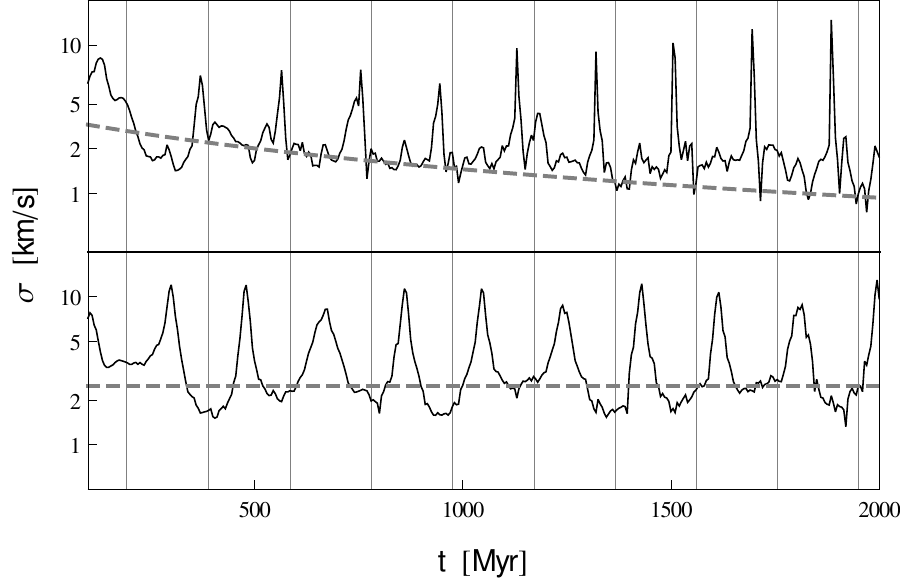}
\caption{\textit{Evolution of the velocity dispersion for a high eccentricity orbit ($b=0.75$). The dispersion parallel to the stream is plotted at the top, the one perpendicular to the stream at the bottom. The grey dahsed curves show the time evolution predicted by the model. The vertical lines correspond to the apocenter passage of the cluster.}}
\label{sigma}
\end{figure}
Fig. \ref{sigma} can be understood qualitatively by assuming that the particles in the stream are on nearly free epicyclic orbits around the host, which means that the host potential is dominating and that the stream particles are not feeling each other. Slightly displaced orbits are then crossing at apo- and again at pericenter which leads to large peaks in the velocity dispersion.

Our model predicts a longitudinal dispersion that decreases on average, an effect that is not clearly visible in the plot on the top of Fig. \ref{sigma}. Whilst the minima in the longitudinal dispersion seem to decrease as predicted, the maxima are growing with time. This growth comes from the fact that the particle orbits separate more and more to end up at distinct free orbits with the same eccentricity but with a shift in the azimuthal angle. The particles are then all crossing at the same place leading to a sharp peak in the dispersion. The orbital oscillation is also visible in the plot of the transversal dispersion at the bottom of Fig. \ref{sigma}. On average however the transversal dispersion seems to stay constant as predicted by the model. A more detailed study of the stream dispersion was done by \citet{Helmi1999}, who found a similar evolution of the dispersion over many more orbital periods.

\end{list}

\section{Conclusions}
We have studied the gravitational stability of tidal streams by modelling them as thin linearly expanding cylinders of collisionless matter. Such a model leads to a stability criterion that has an exponential dependence on the one dimensional Toomre parameter. 
We derive a perturbation analysis and also use energetic arguments, to show that a cylinder with the dispersion, the density and the growth rate of a tidal stream is stable for all times. 

We used numerical simulations to test our main approximations and to study the detailed phase space evolution of tidal streams. As a final consistency check, we note that none of our simulations show any evidence for gravitational instability.

In reality, a stream is only linearly expanding on average, its length is oscillating during one orbit. This leads to a time interval between apo- and pericenter, where the scale factor shrinks again and the stream is in an unstable regime. Nevertheless, this time interval is too short for the perturbations to grow substantially and the oscillation of the stream length has therefore no influence on the stability.

Collisionless stellar or dark matter streams should therefore evolve smoothly in time, simply stretching further away from the parent system. The structure observed in tidal streams, such as Palomar 5 must have an external origin, perhaps disk shocking or encounters with molecular clouds or dark matter substructures.

Our stability analysis could in principle also be extended to other systems producing streams. However, systems with non spherical shapes and net angular momentum are extremely difficult to analyse with analytical methods, since the alignment of the interacting objects is important. Merging disk galaxies for example produce streams with internal structures strongly depending on the initial alignment of the disks and on their angular momenta. In such cases, high resolution numerical simulations are the indispensable tool for a consistent stability analysis.

\section*{Acknowledgements}
We thank Sebastian Elser and George Lake for helpful discussions. This research is supported by the Swiss National Foundation.

{}

\label{lastpage}
\end{document}